\documentclass[aps,prb,reprint,amsmath,longbibliography]{revtex4-2}
\usepackage{bbm}
\usepackage{mathrsfs}
\usepackage{amsmath}
\usepackage{amsfonts}
\usepackage[colorlinks=true, 
            citecolor=blue, 
            anchorcolor=blue, 
            linkcolor=blue, 
            urlcolor=blue]{hyperref}
\usepackage{graphicx,epstopdf}
\usepackage{subfigure}
\usepackage{epsfig}
\usepackage{dcolumn}
\usepackage{bm}
\usepackage{color}
\usepackage{natbib}
\usepackage{amssymb} 
\usepackage{xcolor}
\usepackage{braket}
\definecolor{mycolor}{HTML}{0681F3}
\begin{document}

\title{Symmetry-driven anisotropic coupling effect in antiferromagnetic topological insulator: Mechanism for high-Chern-number quantum anomalous Hall state}
\author{Yiliang Fan$^{1}$, Huaiqiang Wang$^{2,\ast}$, Peizhe Tang$^{3,4}$, Shuichi Murakami$^{5}$, Xiangang Wan$^{1,6}$, Haijun Zhang$^{1,6,\ast}$, and Dingyu Xing$^{1,6}$}

\affiliation{
 $^1$ National Laboratory of Solid State Microstructures and School of Physics, Nanjing University, Nanjing 210093, China\\
 $^2$ School of Physics and Technology, Nanjing Normal University, Nanjing 210023, China\\
 $^3$ School of Materials Science and Engineering, Beihang University, Beijing 100191, China\\
 $^4$ Max Planck Institute for the Structure and Dynamics of Matter, Center for Free Electron Laser Science, Hamburg 22761, Germany\\
 $^5$ Department of Physics, Tokyo Institute of Technology, Tokyo 152-8551, Japan\\
 $^6$ Collaborative Innovation Center of Advanced Microstructures, Nanjing University, Nanjing 210093, China\\
}

\begin{abstract}
Antiferromagnetic (AFM) topological insulators (TIs), which host magnetically gapped Dirac-cone surface states and exhibit many exotic physical phenomena, have attracted great attention. Here, we find that the coupled surface states can be intertwined to give birth to a set of $2n$ unique new Dirac cones, dubbed intertwined Dirac cones, through the anisotropic coupling enforced by crystalline $n$-fold  ($n=2, 3, 4, 6$) rotation symmetry $C_{nz}$ in the presence of a $PT$-symmetry breaking potential, for example, an electric field. Interestingly, we also find that the warping effect further drives the intertwined Dirac-cone state into a quantum anomalous Hall phase with a high Chern number ($C=n$). Then, based on first-principles calculations, we have explicitly demonstrated six intertwined Dirac cones and a Chern insulating phase with a high Chern number ($C=3$) in MnBi$_2$Te$_4$$/$(Bi$_2$Te$_3$)$_{\mathrm{m}}/$MnBi$_2$Te$_4$ heterostructures, as well as the $C=2$ and $C=4$ phases in HgS and $\alpha$-Ag$_2$Te films, respectively. This work discovers the intertwined Dirac-cone state in AFM TI thin films, which reveals a mechanism for designing the quantum anomalous Hall state with a high Chern number and also paves a way for studying highly tunable high-Chen-number flat bands of twistronics. \end{abstract}

\email{zhanghj@nju.edu.cn}
\email{hqwang@njnu.edu.cn}

\maketitle

\section{Introduction}

The interplay between magnetism and topology in condensed matters has greatly enriched the research content of topological quantum physics with many exotic physical phenomena, such as quantum anomalous Hall effect (QAHE)~\cite{Haldane1988,Yu2010quantized,Chang2013science,Deng2020,Chang2023rmp}, magnetic Weyl semimetals~\cite{Wan2011Weyl,Xu2011Weyl}, topological magnetoelectric effects~\cite{Qi2008Topological,wu2016quantized}, axion polariton~\cite{li2010dynamical, nenno2020axion, Xiao2021Nonlinear,Zhu2022Axionic} and so on. Recently, MnBi$_2$Te$_4$ and its family were discovered to be an important class of promising intrinsic magnetic topological insulators (TIs) and have attracted great attention~\cite{Gong2019cpl, Otrokov2019nature, Zhang2019mbt, Li2019sa, Chen2019intrinsic, chen2019prx-ARPES,li2019prx-ARPES,hao2019prx-ARPES,Otrokov2019prl, Sun2019rational,Ge2020high, Hu2020van,Wang2020dynamical, Fu2020exchange,Lian2020prl,Sass2020prl,Liu2020robust,Gu2021spectral, Li2021prl,gao2021layer,  liu2021magnetic, Zhu2023Floquet,Bai2023Quantized,Wang2023nature,Gao2023nonlinear, zhu2021tunable}. Interestingly, the layer Hall effect was demonstrated in even-septuple-layer (SL) MnBi$_2$Te$_4$ films hosting two separated magnetically gapped Dirac cones exhibiting half-integer quantized Hall conductivity with opposite signs on the top and bottom surfaces~\cite{gao2021layer}. The coupling between these two topological surface states play a crucial role in the low-energy physics especially for a thin TI film. Generally speaking, due to the $n$-fold ($n=2,3,4,6$) crystalline rotation symmetry ($C_{nz}$) that a (magnetic) TI usually respects, the surface-state coupling should contain both isotropic terms \cite{Shan2010, Lu2010Massive, Sun2020prb, Lei2020pnas,Wang2023Dirac,wang2023three} and symmetry-enforced anisotropic terms~\cite{Liu2010prb}. However, to our knowledge, the latter is significantly underrated in the literature and has not been fully appreciated before.


\begin{figure*}[htbp]
\centering
\includegraphics[width=7in]{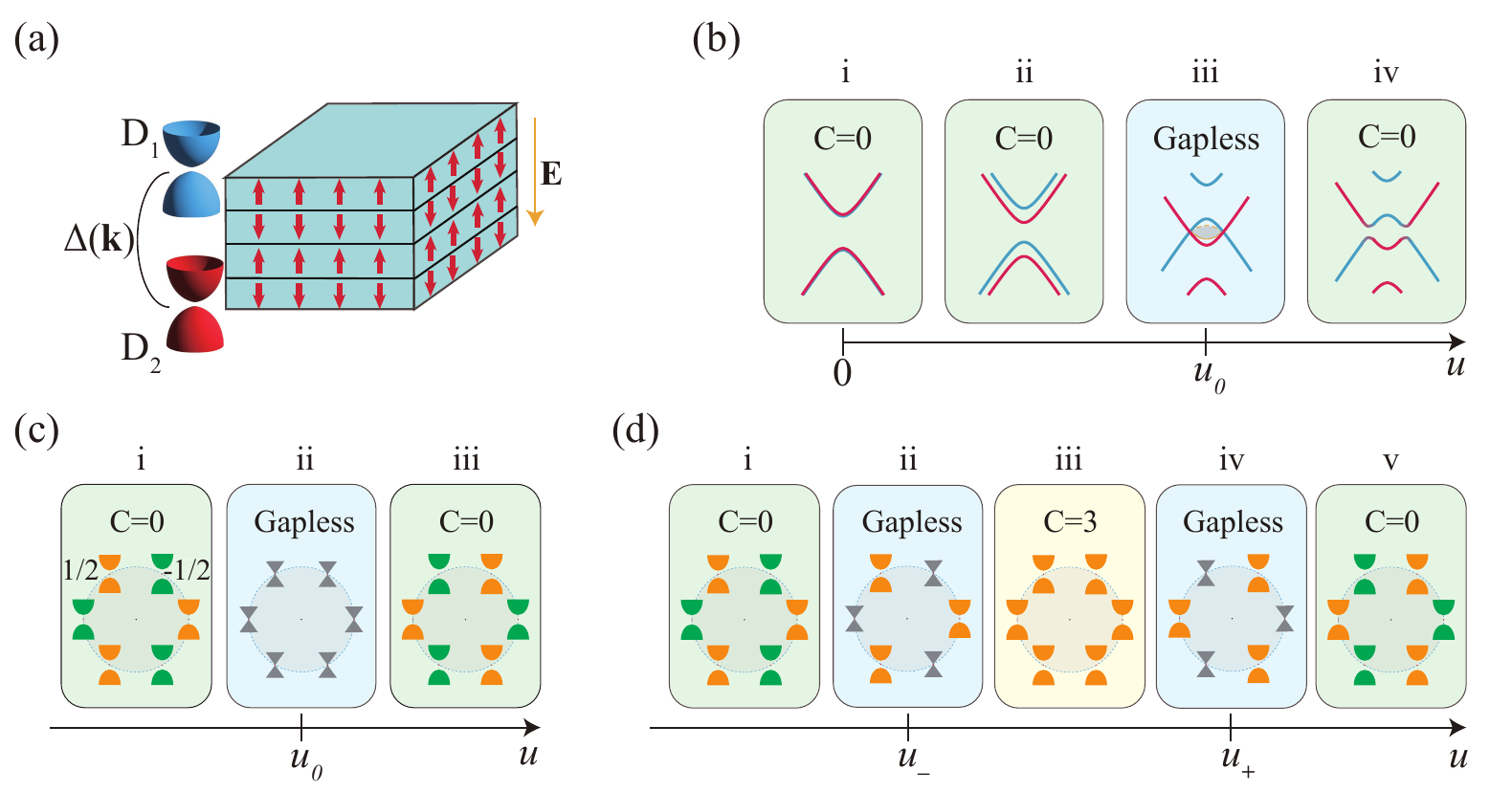}
\caption{ (a) Schematic of the AFM TI thin film with coupled top and bottom Dirac-cone surface states. (b) The surface states are doubly degenerate when the $PT$ symmetry is preserved (i). A $PT$-breaking potential $u$ from the electric field can shift the surface Dirac cones to split the degenerate bands (ii). When only considering isotropic couplings, the energy gap first closes along a nodal ring at $u=u_0$ (iii) and reopens (iv), but without a topological transition. For an AFM TI film with $C_{3z}$ symmetry, (c) the emergence of six intertwined Dirac cones from splitting the nodal ring by the anisotropic coupling. The color of Dirac cones indicates the different Chern number, with green (orange) representing $C=-1/2$ ($C=1/2$). (d) The emergence of the high-Chern-number phase with $C=3$ when further considering the symmetry-enforced warping effect, where the green (orange) intertwined Dirac cone changes its Chern number by 1 ($-$1) at the transition point of $u=u_-$ ($u=u_+$).}
\label{fig1}
\end{figure*}

Remarkably, in this work, we find that the $C_{nz}$-symmetry-enforced anisotropic coupling can induce a set of $2n$ new Dirac cones, termed intertwined Dirac cones, around the $\Gamma$ point through intertwining the two original Dirac cones on the top and bottom surfaces of an antiferromagnetic (AFM) TI thin film under a $PT$-symmetry breaking potential, e.g. an out-of-plane electric field. More intriguingly, when further considering the symmetry-enforced warping effect, an electrically tunable high-Chern-number ($|C|=n$) QAHE~\cite{Bosnar2023,Han2023,Zhu2022,Wan2024} can appear. These are further explicitly verified by the existence of six intertwined Dirac cones and corresponding $|C|=3$ Chern insulating phase in MnBi$_2$Te$_4$$/$(Bi$_2$Te$_3$)$_{\mathrm{m}}/$MnBi$_2$Te$_4$ ($\mathrm{m}=$0, 1, 2) heterostructures through first-principles calculations, and also confirmed in HgS~\cite{Fran2011HgS} and $\alpha$-Ag$_2$Te \cite{Zhang2011Ag2Te} films with $C_{2z}$ and $C_{4z}$ symmetries, respectively, via magnetic proximity effect on the surfaces. Our work not only reveals a concept of intertwined Dirac cones, but also provides a scheme to achieve high-Chern-number QAHE and paves a way to study highly tunable high-Chen-number flat bands of twistronics in AFM TI films.

\section{Methods}

The first-principles calculations are carried out using the Perdew-Burke-Ernzerhof-type (PBE) generalized gradient approximation (GGA)~\cite{GGA1999} of the density functional theory (DFT), using the Vienna \emph{ab initio} simulation package (VASP)~\cite{vasp_prb1996}. The spin-orbit coupling is included in the DFT calculations. We take the GGA + U method with U = 5.0 eV to investigate the correlation effects. For the three materials:  MnBi$_2$Te$_4$$/$(Bi$_2$Te$_3$)$_{\mathrm{m}}/$MnBi$_2$Te$_4$ heterostructures, HgS and $\alpha$-Ag$_2$Te, we use a kinetic energy cutoff of 410 eV, 390 eV, and 380 eV, and a k-point mash of 11 × 11 × 1, 9 × 9 × 9, and 10 × 10 × 8, respectively. We use the experimental lattice constants of MnBi$_4$Te$_7$ ($a = 4.355\ \mathrm{\AA}$ and $c = 23.815\ \mathrm{\AA}$)~\cite{Ziya2019_147sy} to construct the MnBi$_2$Te$_4$$/$(Bi$_2$Te$_3$)$_{\mathrm{m}}/$MnBi$_2$Te$_4$ heterostructures. These heterostructures are fully optimized until the force on each atom is less than 0.01 eV/$\mathrm{\AA}$. The optimized lattice constants are $a=4.274\ \mathrm{\AA}$ (m=0), $a=4.301\ \mathrm{\AA}$ (m=1) and $a=4.311\ \mathrm{\AA}$ (m=2). Because they are two-dimensional heterostructures, we just list the in-plane lattice constants here. The lattice constants of HgS ($a = 5.85\ \mathrm{\AA}$) and $\alpha$-Ag$_2$Te ($a = 6.8\ \mathrm{\AA}$) are obtained from previous works~\cite{Fran2011HgS,Zhang2011Ag2Te}. To simulate the uniaxial compressive strain applied to $\alpha$-Ag$_2$Te, we elongate the in-plane lattice parameter "$a$" and compress the out-of-plane parameter "$c$," while keeping the cell volume unchanged. The same method has been employed in Ref.~\cite{half-heusler2010,ruan2016}. As a result, we obtain $a=6.89\ \mathrm{\AA}$ and $c=6.67\ \mathrm{\AA}$, resulting in a $c$/$a$ ratio of 0.97.

Based on the results of the DFT calculations, we construct the maximally localized Wannier functions (MLWFs)~\cite{wannier90_1997,wannier90_2001} using the WANNIER90 package~\cite{wannier90_2020}. During the construction of the MLWFs, we select the following orbitals: Mn-d, Bi-p, and Te-p for the MnBi$_2$Te$_4$$/$(Bi$_2$Te$_3$)$_{\mathrm{m}}/$MnBi$_2$Te$_4$ heterostructures; Hg-s, Hg-d, and S-p for HgS; and Ag-s, Ag-d, and Te-p for $\alpha$-Ag$_2$Te.

We obtain the Wannier-based tight-binding Hamiltonian after constructing the MLWFs. We utilize the WannierTools package~\cite{wanniertools} to construct the $N$-cell-slab tight-binding Hamiltonians for HgS and $\alpha$-Ag$_2$Te, incorporating magnetic proximity effects and electric fields based on their respective bulk tight-binding Hamiltonians. To simulate the magnetic proximity effect, we add opposite Zeeman coupling energies of 0.05 eV and $-$0.05 eV to the top and bottom cells, respectively, of the $N$-cell-slab tight-binding Hamiltonian. To simulate the electric field, we add an onsite potential to each orbital. The value of this potential equals the \emph{z} coordinate of the Wannier center of the orbital multiplied by the value of the electric field. The calculations of the topological properties of the three materials are also conducted using the WannierTools package.

\section{Intertwined Dirac-cone states}

We start from an AFM TI thin film with two Dirac-cone surface states located on the top and bottom surfaces, respectively, as illustrated in Fig.~\ref{fig1}(a). Due to the out-of-plane surface magnetic moments, each of them can be described by $H_{D_\alpha}=s\hbar v(k_x\sigma_y-k_y\sigma_x)+m_\alpha\sigma_z$, where $\alpha=1,2$, $D_1$ ($D_2$) represents the top (bottom) Dirac cone, $v$ is the Fermi velocity, $s=+(-)$ for the Dirac cone $D_1$ ($D_2$), the Planck's constant $\hbar$ is simply set as 1 henceforth, and $m_\alpha$ indicates Zeeman coupling. Since magnetic moments are opposite on the two surfaces, we have $m_1=-m_2= m$ ($m$ is set to be positive henceforth). In the basis of $|D_1,\uparrow\rangle,|D_1,\downarrow\rangle,|D_2,\uparrow\rangle,|D_2,\downarrow\rangle$, where $\uparrow (\downarrow)$ represents the spin, a four-band low-energy Hamiltonian of the two uncoupled Dirac cones is directly written as
\begin{equation}
    H_{0}= v\tau_z\otimes(k_x\sigma_y-k_y\sigma_x)+m\tau_z\otimes\sigma_z.
\end{equation}
Here, the Pauli matrix $\tau_z$ acts in the subspace of top and bottom Dirac cones. This Hamiltonian describes two degenerate Dirac cones with a gap of $2m$, as shown in Fig.~\ref{fig1}(b)(i). Although the time-reversal ($T=i\sigma_y K$ with $K$ denoting the complex conjugate) and inversion ($P=\tau_x$) symmetries are separately broken, the combined $PT$ symmetry is preserved, which ensures the double degeneracy of each band. We then introduce a $PT$-symmetry breaking potential $u$ between the two Dirac cones described by $H_{u}=u\tau_z\otimes\sigma_0$, which could be induced by an out-of-plane electric field ($u\propto E$). This potential lifts the double degeneracy of each band by conversely shifting the two spatially separated Dirac cones in energy [see Fig. \ref{fig1}(b)(ii)]. With increasing $u$, the energy gap gradually reduces until it closes at $\Gamma$ when $u=m$. Further increase of $u$, i.e., $u>m$, will expand the closing point into a nodal ring.

\begin{figure*}[htbp]
\centering
\includegraphics[width=7in]{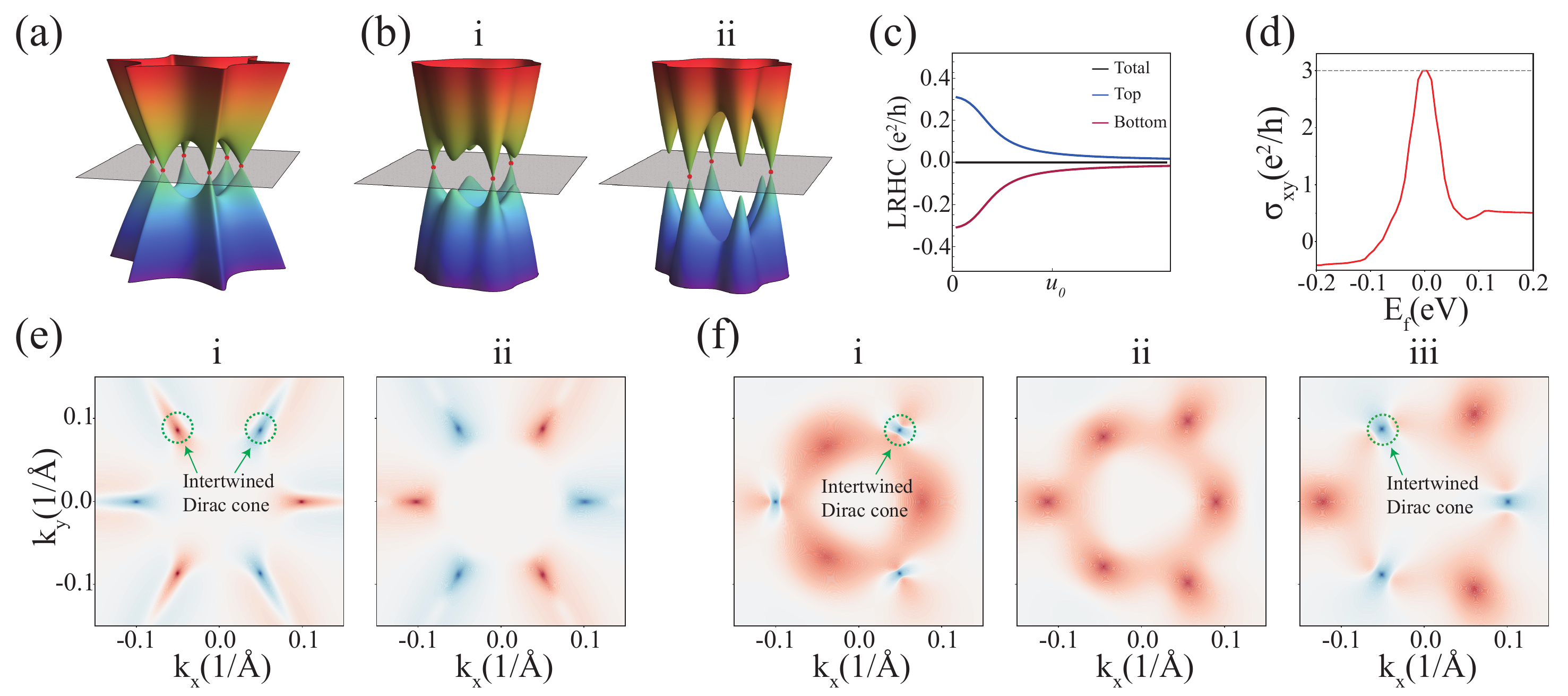}
\caption{The band structures for (a) the six massless intertwined Dirac cones at the critical potential $u=u_0$ without the warping effect, and (b) the two sets of three massless intertwined Dirac cones at the gap-closing point of $u=u_-$ (i) and $u=u_+$ (ii). (c) The layer-resolved Hall conductivity (LRHC) for the top (blue line) and bottom (red line) surfaces. (d) The anomalous Hall conductivity (AHC) for the $u_-<u<u_+$ case, where the AHC equals $3\ e^2/h$ when the Fermi level lies in the energy gap. (e) Without the warping effect, the Berry curvatures for $u<u_0$ (i) and $u>u_0$ (ii), respectively. (f) With the warping effect, typical Berry curvatures for $u<u_-$ ($C=0$) (i), $u_-<u<u_+$ ($C=3$) (ii), and $u>u_+$ ($C=0$) (iii), respectively. The parameters are chosen as $m=0.05\ \mathrm{eV}, v=1\ \mathrm{eV\cdot \AA}, \Delta=0.05\ \mathrm{eV},B=5\ \mathrm{eV\cdot \AA^2}, R_1=100\ \mathrm{eV\cdot\AA^3}$, and $R_2=100\ \mathrm{eV\cdot\AA^3}$ in the numerical calculations.}
\label{fig2}
\end{figure*}

Now we consider the coupling between $D_1$ and $D_2$.  When assuming that the AFM TI respects $C_{nz}$ ($n=2,3,4,6$) rotation symmetry and a combined symmetry $M_xT$ of mirror and time-reversal operations, we can obtain the following coupling Hamiltonian [see the Supplemental Material (SM)~\cite{SM} for more details]:
\begin{equation}
\label{Hn}
    H_{\mbox{\tiny coup}}=(\Delta-Bk^2)\tau_x\otimes \sigma_0+R_1g_n(k_x,k_y)\tau_y\otimes \sigma_0,
\end{equation}
where $k^2=k_x^2+k_y^2$ and $g_n(k_x,k_y)=(k_+^n-k_-^n)/2i$ with $k_{\pm}=k_x\pm ik_y$, and $i$ is the imaginary unit. The first term in Eq.~(\ref{Hn}) describes isotropic couplings up to $k^2$ order ($\Delta B>0$ is assumed throughout the work~\cite{Lu2010Massive}), while the second term comes from the symmetry-enforced anisotropic coupling (for simplicity we have dropped other higher-order coupling terms which will not affect our main results~\cite{SM}). Most importantly, it plays a crucial role in generating intertwined Dirac-cone states, as shown below. Without the anisotropic coupling, the energy gap closes along a nodal ring with the radius $k_0=\sqrt{\Delta/B}$ at $u=u_0=\sqrt{m^2+v^2\Delta/B}$ [see Fig.~\ref{fig1}(b)(iii)], and it reopens when $u>u_0$ [see Fig.~\ref{fig1}(b)(iv)]. However, there is no topological phase transition in this gap-closing-and-reopening process, for the Chern number remains zero. Interestingly, due to the anisotropic coupling, the nodal ring at $u=u_0$ splits into $2n$ nodal points connected to each other by $C_{nz}$ symmetry, as illustrated in Fig.~\ref{fig1}(c)(ii). In the $k$-$\theta$ polar coordinate, these nodal points are located at $(k_0,\theta_{j})$, where $\theta_j=j\pi/n$, with $j=0,1,2,...,2n-1$. Only at these nodal points does the coupling become zero, and by expanding the Hamiltonian around them to linear order, we obtain a low-energy Dirac Hamiltonian as 
\begin{equation}
\label{heff}
h_n(\theta_{j})=-2Bk_0q_{\rho}\sigma_x+(-1)^{j}nR_1k_0^{n-1}q_\theta\sigma_y+\delta u\sigma_z,
\end{equation}
where $q_\rho=k-k_0$ and $q_\theta=k_0\delta\theta$ with $\delta\theta=\theta-\theta_{j}$, are the momenta measured from each nodal point along the radial and angular directions in the polar coordinates  and $\delta u=u-u_0$. The effective Hamiltonian $h_n(\theta_{j}) $ indicates that each nodal point is a new Dirac-cone state with a mass of $\delta u$. Since this set of Dirac-cone states emerge from the coupling between the top and bottom surface states, they are dubbed "intertwined" Dirac-cone states. Moreover, according to Eq.~(\ref{heff}), the helicity of the intertwined Dirac-cone state labeled by odd or even value of $j$ is opposite. Consequently, its Chern number, given by $C(\theta_{j})=\frac{(-1)^{j+1}}{2}{\rm sgn}(\delta u)$, becomes opposite for odd and even $j$ [see Fig.~\ref{fig1}(c)(i)]. When $\delta u$ changes from negative ($u<u_0$) to positive ($u>u_0$), $C(\theta_{j})$ changes by $+1$ $(-1)$ for odd (even) $j$ [see Fig.~\ref{fig1}(c)(iii)], so the total Chern number still equals zero. To exemplify this, we have chosen $n=3$ to numerically calculate typical Berry curvature distributions of the valence bands near the transition point with $u<u_0$ [Fig.~\ref{fig2}(e)(i)] and $u>u_0$ [Fig.~\ref{fig2}(e)(ii)], as well as the band structure of the massless intertwined Dirac-cone states at the transition point of $u=u_0$ [Fig.~\ref{fig2}(a)]. Obviously, the local Berry curvature around each intertwined Dirac-cone state changes its sign between $u<u_0$ and $u>u_0$, and it is always opposite between odd and even $j$, leading to a zero net Chern number.

Furthermore, we have also investigated how the layer Hall effect is affected by the coupling between surface states. In Fig.~\ref{fig2}(c), we have plotted the layer-resolved Hall conductivity (LRHC) versus the potential $u$. With increasing $u$, the LRHCs of both surfaces gradually decrease toward zero. Note that the deviation from the quantized value $\pm e^2/2h$ before the gap closing $(u<u_0)$ results from the coupling between the surface states.

\begin{figure*}[htbp]
\centering
\includegraphics[width=7in]{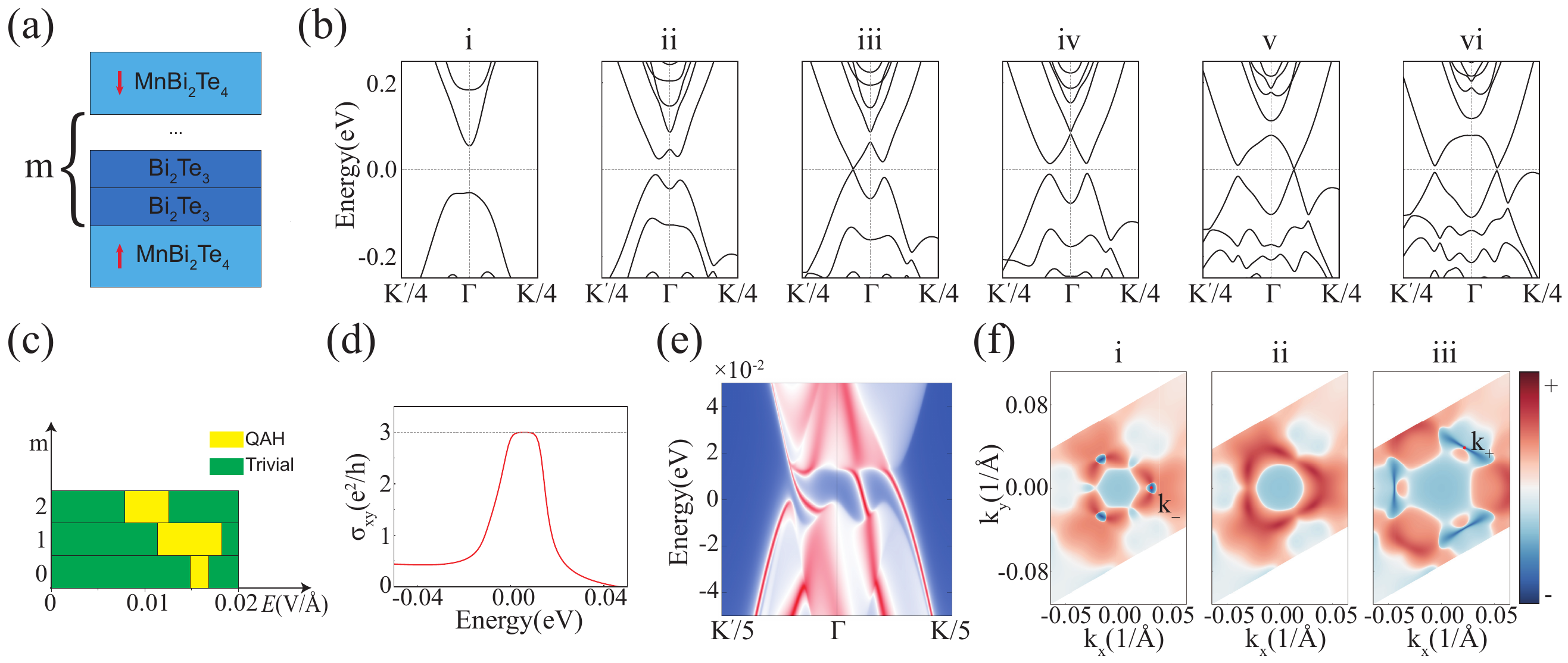}
\caption{(a) Schematic of the MnBi$_2$Te$_4$ $/$(Bi$_2$Te$_3$)$_{\mathrm{m}}$$/$MnBi$_2$Te$_4$ ($\mathrm{m}=$0, 1, 2) heterostructure structure. (b) The band structures for $\mathrm{m}=1$ with increasing the electric field, where $E=0.0\ {\rm V/\AA}$ (i), $0.008\ {\rm V/\AA}$ (ii), $0.0113\ {\rm V/\AA}$ (iii), $0.0135\ {\rm V/\AA}$ (iv), $0.0182\ {\rm V/\AA}$ (v), and $0.02\ {\rm V/\AA}$ (vi). (c) The topological phase diagram for different values of m, where the yellow (green) region denotes the topologically nontrivial (trivial) phase with $C=3$ ($C=0$). (d), (e) The AHC and chiral edge states for the $C=3$ phase at $E=0.0135\ {\rm V/\AA}$ for $\mathrm{m}=1$. A quantized value of $\sigma_{xy}=3\ e^2/h$ can be seen in (d) when the Fermi level is in the energy gap, which originates from the emergence of three chiral edge states shown in (e). (f) Berry curvatures for $E=0.011\ {\rm V/\AA}$ ($u<u_-$ with $C=0$) (i),  $E=0.0135\ {\rm V/\AA}$ ($u_-<u<u_+$ with $C=3$) (ii), and $E=0.0186\ {\rm V/\AA}$ ($u>u_+$ with $C=0$) (iii). Significant changes of Berry curvatures are seen around the intertwined Dirac cones located at $k_-$ ($k_+$) in the first (second) topological phase transition.}
\label{fig3}
\end{figure*}

\section{ Warping effect and high-Chern-number phase}

Based on the above analysis, we now reveal the emergence of a high-Chern-number phase with $|C|=n$ by further considering the symmetry-enforced warping effect~\cite{Fu2009warping, Naselli2022Magnetic} of the two original Dirac-cone surface states. The warping effect can be described by the Hamiltonian $H_{\mbox {\tiny warp}}=R_2w_n(k_x,k_y)\tau_z\otimes \sigma_z$, where $w_n(k_x,k_y)=(k_+^n+k_-^n)/2$. Because of $H_{\mbox {\tiny warp}}$, the critical potential $u_0$ of the gap-closing condition for each intertwined Dirac-cone state now becomes \cite{SM}
\begin{equation}
    u_{0,j} =\sqrt{\Big [ m+(-1)^j R_2\left(\Delta/B\right)^{n/2}\Big ]^2+v^2\Delta/B}.
\end{equation}
In contrast to the case without the warping effect where $u_{0,j}$ is identical for all values of $j$, the critical gapless transition point is now distinct between odd and even $j$, namely, $u_{0,j}\equiv$$u_-$ ($u_+$) for odd (even) $j$. This can be understood from the fact that $H_{\mbox {\tiny warp}}$ effectively introduces opposite corrections to the Zeeman term between intertwined Dirac cones with odd and even $j$. The difference between $u_{\pm}$ indicates that the changes of Chern number $\Delta C$ of intertwined Dirac-cone states with odd $j$ ($\Delta C=+1$) and even $j$ ($\Delta C=-1$) no longer occur simultaneously [see Figs.~\ref{fig1}(d)(ii) and \ref{fig1}(d)(iv)]. Therefore, if $mR_2>0$ is assumed, we have $u_-<u_+$ and a high-Chern-number phase with $C=n$  when $u_-<u<u_+$, while $C=0$ for both $u>u_+$ and $u<u_-$, as exemplified by the $C=3$ case in Fig.~\ref{fig1}(d). We have also calculated the Berry curvature distribution of the valence bands with $u<u_-$ [Fig.~\ref{fig2}(f)(i)], $u_-<u<u_+$  [Fig.~\ref{fig2}(f)(ii)], and $u>u_+$  [Fig.~\ref{fig2}(f)(iii)]. The band structures at the critical transition points of $u=u_-$ and $u=u_+$ are presented in Figs.~\ref{fig2}(b)(i) and \ref{fig2}(b)(ii), respectively. The sign change of the Berry curvature around each intertwined Dirac-cone state across the transition points of $u_{\pm}$ can be clearly seen in Fig.~\ref{fig2}(f). Moreover, to verify the high-Chern-number phase for $u_-<u<u_+$, we have calculated the anomalous Hall conductivity (AHC) in Fig.~\ref{fig2}(d), where the AHC exactly equals $3\ e^2/h$ when the Fermi energy lies in the gap.

\section{Material Realization}

\subsection{A. MnBi$_2$Te$_4$ $/$(Bi$_2$Te$_3$)$_{\mathrm{m}}$$/$MnBi$_2$Te$_4$}

Inspired by recent experimental progress on MnBi$_2$Te$_4$ family intrinsic AFM TIs and related heterostructures, such as MnBi$_4$Te$_7$ and MnBi$_6$Te$_{10}$~\cite{Ziya2019_147sy,Vidal2019topological,chen2019prb, Wu2019natural,souchay2019jmcc,Wu2020Distinct,Hu2020van,vidal2021prl,ge2022direct,xu2022nc}, we take the MnBi$_2$Te$_4$ $/$(Bi$_2$Te$_3$)$_{\mathrm{m}}$$/$MnBi$_2$Te$_4$ ($\mathrm{m}=$0, 1, 2) heterostructure as a realistic example of the preceding $n=3$ case with the $C_{3z}$ symmetry and perform first-principles calculations to demonstrate the existence of six intertwined Dirac-cone states and a high-Chern-number ($C=3$) QAHE. As shown in Fig.~\ref{fig3}(a), the MnBi$_2$Te$_4$$/$(Bi$_2$Te$_3$)$_{\mathrm{m}}/$MnBi$_2$Te$_4$ heterostructure is constructed by inserting $\mathrm{m}$ Bi$_2$Te$_3$ quintuple-layers  between two MnBi$_2$Te$_4$ SLs, which preserves the required $C_{3z}$ and the combined $M_xT$ symmetries. Moreover, according to the total energy calculation~\cite{SM}, the out-of-plane AFM order is found to be the magnetic ground state.

First, we inspect the band structure of the MnBi$_2$Te$_4$$/$Bi$_2$Te$_3$$/$MnBi$_2$Te$_4$ heterostructure with increasing the out-of-plane electric field. Without the electric field, the band structure is doubly degenerate with an energy gap [see Fig.~\ref{fig3}(b)(i)]. Once a weak electric field is applied (e.g. $E = 0.008$ V/\AA), the band structure starts splitting due to breaking the $PT$ symmetry, seen in Fig.~\ref{fig3}(b)(ii). With increasing the electric field, the band structure undergoes two successive gap-closing-and-reopening processes [see Figs.~\ref{fig3}(b)(iii) and (v)], where the gap closes at $E = 0.0113$ V/\AA \ and $E = 0.0182$ V/\AA, respectively. Next, we investigate topological properties, namely, Berry curvatures and corresponding Chern numbers for different regions of the electric field. The Chern number is calculated to be $C=3$ in the intermediate region between the two gap-closing processes, whereas $C=0$ for the other two regions (see the Wilson-loop~\cite{wilson-loop} calculations in the SM \cite{SM}), which is well consistent with the above model prediction. Moreover, we have calculated the AHC in the $C=3$ region [see Fig.~\ref{fig3}(d)], which takes the expected value of $\sigma_{xy}=3\ e^2/h$ when the Fermi level lies in the gap. This is further confirmed by the emergence of three chiral edge states in the energy gap [see Fig.~\ref{fig3}(e)]. Typical Berry curvatures for the three regions are shown in Fig.~\ref{fig3}(f), where significant changes of the Berry curvature occur around the points of intertwined Dirac cones, reflecting their pivotal role in realizing the high-Chern-number QAHE. Similar topological phase diagrams can be found for other values of m, as shown in Fig.~\ref{fig3}(c) \cite{SM}.

\begin{figure*}[htbp]
\centering
\includegraphics[width=7in]{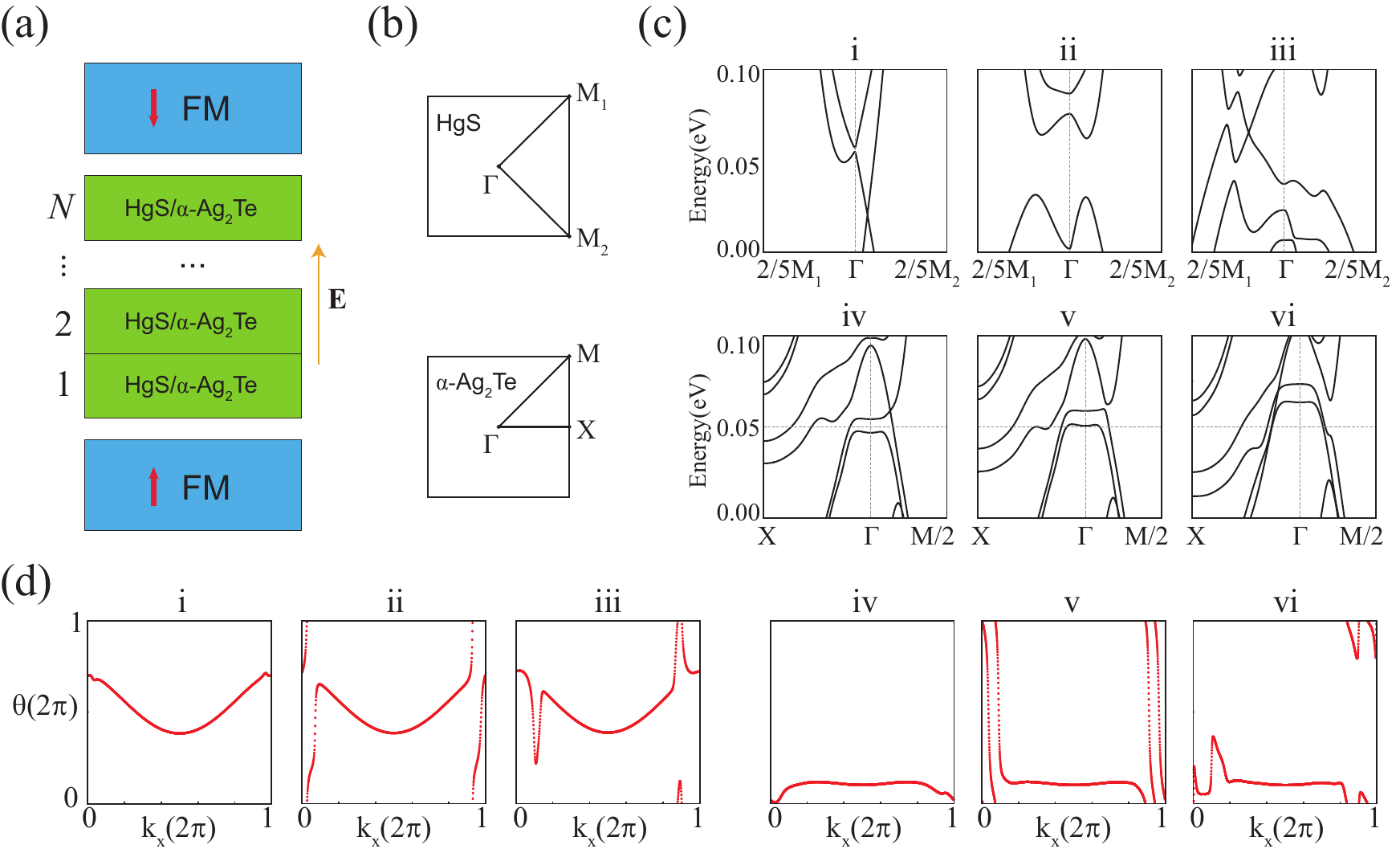}
\caption{(a) Schematic of the $N$-cell slab of HgS or $\alpha$-Ag$_2$Te. The top and bottom surfaces of the slab are in proximity to ferromagnets with opposite magnetic moments. (b) The 2D Brillouin zone and high-symmetry points of HgS and $\alpha$-Ag$_2$Te. (c) Band structures for HgS and $\alpha$-Ag$_2$Te with increasing the electric field, where $E=0.0233\ {\rm V/\AA}$ (i), $0.03\ {\rm V/\AA}$ (ii), and $0.0384\ {\rm V/\AA}$ (iii) for HgS and $0.027\ {\rm V/\AA}$ (iv), $0.03\ {\rm V/\AA}$ (v), and $0.039\ {\rm V/\AA}$ (vi) for $\alpha$-Ag$_2$Te. (d) Wilson-loops of HgS and $\alpha$-Ag$_2$Te with increasing the electric field, where $E=0.02\ {\rm V/\AA}$ (i), $0.03\ {\rm V/\AA}$ (ii), and $0.04\ {\rm V/\AA}$ (iii) for HgS and $0.014\ {\rm V/\AA}$ (iv), $0.031\ {\rm V/\AA}$ (v), and $0.041\ {\rm V/\AA}$ (vi) for $\alpha$-Ag$_2$Te.}
\label{fig4}
\end{figure*}

\subsection{B. HgS and $\alpha$-Ag$_2$Te}
We choose $\mathrm{HgS}$ and $\alpha$-$\mathrm{Ag_2Te}$ as realistic materials to realize the models for $n=2$ and $n=4$, respectively. HgS is a topological insulator respecting $C_{2z}$ symmetry~\cite{Fran2011HgS}. $\alpha$-$\mathrm{Ag_2Te}$ is a zero-gap semiconductor respecting $C_{4z}$ symmetry with inverted band structure~\cite{Zhang2011Ag2Te}. Thus, we apply a uniaxial compressive strain along the "$c$" axis, resulting in $c/a=0.97$ ($c/a=1$ before applying strain), to open a small gap at the $\Gamma$ point. Moreover, since $\mathrm{HgS}$ and $\alpha$-$\mathrm{Ag_2Te}$ are nonmagnetic, for each material, we have constructed a $N$-cell-slab tight-binding model along the out-of-plane direction based on the first-principles calculations of the bulk electronic structures and suppose there are two ferromagnetic materials with opposite magnetic moments near the top and bottom surfaces of the slab separately. To simulate the magnetic proximity effect induced by the two ferromagnetic materials, we add opposite Zeeman coupling energies to the top ($0.05\ \mathrm{eV}$) and bottom ($-0.05\ \mathrm{eV}$) cells, respectively. Finally, the effect of the out-of-plane electric field is simulated by adding an onsite potential to each orbital. The value of the potential equals the coordinates of the Wannier center of the orbital along the stacking direction multiplied by the value of the electric field.

For HgS and $\alpha$-$\mathrm{Ag_2Te}$, we have built a five-cell-slab and four-cell-slab tight-binding model, respectively, as shown in Fig. \ref{fig4}(a). Figure \ref{fig4}(c) shows the band structures of the model under different electric fields, where the high-symmetry points have been labeled in Fig. \ref{fig4}(b). We find that the energy gap for HgS closes at $E=0.0233\ \mathrm{V/\AA}$ [Fig. \ref{fig4}(c)(i)] and $0.0384\ \mathrm{V/\AA}$ [Fig. \ref{fig4}(c)(iii)], while for $\alpha$-$\mathrm{Ag_2Te}$, it closes at $E=0.027\ \mathrm{V/\AA}$ [Fig. \ref{fig4}(c)(iv)] and $0.039\ \mathrm{V/\AA}$ [Fig. \ref{fig4}(c)(vi)], respectively. Between the two electric field values, there is a $|C|=2$ phase for HgS and $|C|=4$ phase for $\alpha$-$\mathrm{Ag_2Te}$, which is verified by the calculations of the Wilson-loops in Fig. \ref{fig4}(d)(i)-(iii) (HgS) and Fig. \ref{fig4}(d)(iv)-(vi) ($\alpha$-Ag$_2$Te). Figure \ref{fig4}(d)(ii) and Fig. \ref{fig4}(d)(v) clearly shows the existence of the $|C|=2$ and $|C|=4$ phase.

\section{Conclusion}

In summary, we have considered the previously overlooked anisotropic coupling between Dirac-cone surface states of AFM TI thin films with $n$-fold rotational symmetry. Intriguingly, this coupling could lead to the emergence of $2n$ intertwined Dirac cones under an out-of-plane electric field, and a high-Chern-number phase with $|C|=n$ is predicted by tuning the electric field, which has been explicitly verified by first-principles calculations. The proposed intertwined Dirac-cone states are essentially different from conventional topological Dirac-cone surface states. The most unique advantage of the intertwined Dirac cones is the flexible electrical tunability, which could give rise to interesting physical phenomena. For example, a relative twist between the top and bottom surfaces of the AFM TI could result in flat bands with significantly reduced velocity of the intertwined Dirac-cone state, which may lead to high-Chern-number flat bands. More interestingly, the flat bands are expected to be greatly tuned by an electric field, which provides a new versatile platform for studying the interplay between topology, magnetism, twistronics, and strong correlation effects. 

\begin{acknowledgments}
This work is partly supported by National Key Projects for Research and Development of China (Grants No. 2021YFA1400400 and No. 2022YFA1403602), the Fundamental Research Funds for the Central Universities (Grant No. 020414380185), the Natural Science Foundation of Jiangsu Province (No. BK20200007), the Natural Science Foundation of China (No. 12074181, No. 12104217, No. 12174182, No. 12234011, No. 12374053, and No. 11834006), and the e-Science Center of Collaborative Innovation Center of Advanced Microstructures. S. M. is supported by JSPS KAKENHI Grants No.~JP22H00108 and No.~JP22K18687 and MEXT Initiative to Establish Next-generation Novel Integrated Circuits Centers (X-NICS) Grant No. JPJ011438.
\end{acknowledgments}

\bibliography{ref}

\end{document}


\title{Supplemental Materials
 for ''Symmetry-driven anisotropic coupling effect in antiferromagnetic topological insulator: Mechanism for high-Chern-number quantum anomalous Hall state''}

\author{Yiliang Fan$^{1}$, Huaiqiang Wang$^{2,\ast}$, Peizhe Tang$^{3,4}$, Shuichi Murakami$^{5}$, Xiangang Wan$^{1,6}$, Haijun Zhang$^{1,6,\ast}$, and Dingyu Xing$^{1,6}$}

\affiliation{
 $^1$ National Laboratory of Solid State Microstructures and School of Physics, Nanjing University, Nanjing 210093, China\\
 $^2$ School of Physics and Technology, Nanjing Normal University, Nanjing 210023, China\\
 $^3$ School of Materials Science and Engineering, Beihang University, Beijing 100191, China\\
 $^4$ Max Planck Institute for the Structure and Dynamics of Matter, Center for Free Electron Laser Science, Hamburg 22761, Germany\\
 $^5$ Department of Physics, Tokyo Institute of Technology, Tokyo 152-8551, Japan\\
 $^6$ Collaborative Innovation Center of Advanced Microstructures, Nanjing University, Nanjing 210093, China\\
}

\email{zhanghj@nju.edu.cn}
\email{hqwang@njnu.edu.cn}
\maketitle

\tableofcontents

\section{Detailed solutions of the intertwined Dirac cones}

Here, we present the detailed solutions of the intertwined Dirac cones induced by the anisotropic coupling through solving the gap-closing points and critical potentials of the model Hamiltonian without/with the warping term. Without the warping term, the total Hamiltonian is given by
\begin{equation}
H_0+H_{\mbox{\tiny{coup}}}=m\tau_z\otimes\sigma_z+u\tau_z\otimes\sigma_0+v\tau_z\otimes(k_x\sigma_y-k_y\sigma_x)+(\Delta-Bk^2)\tau_x\otimes\sigma_0+R_1g_n(k_x,k_y)\tau_y\otimes\sigma_0
\label{eq1}
\tag{S1}
\end{equation}
where $g_n(k_x,k_y)=(k_+^n-k_-^n)/2i$ and $k_{\pm}=k_x\pm ik_y$. To get the gap-closing points and critical potential, we must first solve the eigenvalue of the Hamiltonian, and the result is shown below
\begin{equation}
\tag{S2}
    \epsilon_{\pm,\alpha}=\pm[m^2+v^2k^2+u^2+(\Delta-Bk^2)^2+R_1^2g_n^2+2(-1)^{\alpha-1}u\sqrt{m^2+v^2k^2}]^{1/2}, \ \alpha=1,2
\end{equation}
Next, we choose the two energy eigenvalues near the zero energy, i.e. $\epsilon_{\pm,2}$, and let them equal to each other $\epsilon_{+,2}=\epsilon_{-,2}$, we get the critical potential
\begin{equation}
\tag{S3}
    u_0=\left\{m^2+v^2k^2-(\Delta-Bk^2)^2-R_1^2g_n^2+2i\sqrt{(m^2+v^2k^2)[(\Delta-Bk^2)^2+R_1^2g_n^2]}\right\}^{1/2}
    \label{eq3}
\end{equation}
Since the potential $u_0$ must be real, we naturally have the conditions:
\begin{equation}
\tag{S4}
    \left\{\begin{aligned}
    &\Delta-Bk^2=0\\
    &g_n=(k_+^n-k_-^n)/2i=k^n\sin(n\theta)=0.
    \end{aligned}\right.
\end{equation}
Then we can get the gap-closing points (intertwined Dirac-cone points) from the above conditions and substitute them into the expression of $u_0$ to get the critical potential value. The final results are shown below:
\begin{equation}
\tag{S5}
    \left\{\begin{aligned}
    & u_0=\sqrt{m^2+v^2\Delta/B}\\
    & k_0=\sqrt{\Delta/B}\\
    & \theta_j=j\pi/n,\quad j=0,1,\cdots,2n-1.
    \end{aligned}\right.
\end{equation}

When considering the warping term given by
\begin{equation}
\tag{S6}
    H_{\mbox{\tiny{warp}}}=R_2w_n(k_x,k_y)\tau_z\otimes \sigma_z,
\end{equation}
with $w_n(k_x,k_y)=(k_+^n+k_-^n)/2$, the solutions of the critical potential and gap-closing points can be obtained by substituting $m$ in the above equations with $m(\mathbf{k})=m+R_2w_n(k_x,k_y)$. The results are then obtained as

\begin{equation}
\tag{S7}
    \left\{\begin{aligned}
    & u_{0,j}=\sqrt{\Big [ m+(-1)^j R_2\left(\Delta/B\right)^{n/2}\Big ]^2+v^2\Delta/B}\\
    & k_0=\sqrt{\Delta/B}\\
    & \theta_j=j\pi/n, \quad j=0,1,\cdots,2n-1.
    \end{aligned}\right.
\end{equation}

\section{Low-energy effective Hamiltonians of the intertwined Dirac cones}

First, we need to derive the two eigenstates of each intertwined Dirac cone located at $k=k_0$ and $\theta=\theta_j$ in the polar coordinates, which are given  by

\begin{equation}
\tag{S8}
    \left\{\begin{aligned}
        & \psi_{+}=[\frac{ie^{-i\theta_j}(-m + \sqrt{m^2 + k_0^2 v^2})}{k_0v},1,0,0]^{T}\\
        & \psi_{-}=[0,0,\frac{ie^{-i\theta_j}(-m + \sqrt{m^2 + k_0^2 v^2})}{k_0v},1]^{T}
    \end{aligned}\right.
\end{equation}

Then, by projecting the four-band full Hamiltonian in Eq. (S1) into the subspace expanded by the above two eigenstates, the two-band effective Hamiltonian up to linear order in momentum is then obtained as
\begin{equation}
\tag{S9}
\begin{aligned}
    h(\theta_j)&=\begin{bmatrix}
        \frac{\langle \psi_+|H|\psi_+\rangle}{|\langle\psi_+|\psi_+\rangle|} & \frac{\langle \psi_+|H|\psi_-\rangle}{\sqrt{|\langle\psi_+|\psi_+\rangle||\langle\psi_-|\psi_-\rangle|}}\\
        \frac{\langle \psi_-|H|\psi_+\rangle}{\sqrt{|\langle\psi_+|\psi_+\rangle||\langle\psi_-|\psi_-\rangle|}} & \frac{\langle \psi_-|H|\psi_-\rangle}{|\langle\psi_-|\psi_-\rangle|}
    \end{bmatrix}\\
    &=-2Bk_0q_{\rho}\sigma_x+(-1)^{j}nR_1k_0^{n-1}q_\theta\sigma_y+\delta u\sigma_z,
    \end{aligned}
\label{eq10}
\end{equation}
where $q_\rho=k-k_0$, $q_\theta=k_0\delta\theta$ with $\delta\theta=\theta-\theta_{j}$ are the momentum measured from the nodal points along the radial and angular directions in the polar coordinates  and $\delta u=u-u_0$.  Obviously, this effective Hamiltonian describes a Dirac-cone state.

\section{Discussion of the other neglected anisotropic coupling}

Based on the theory of invariants~\cite{Liu2010prb}, when considering all symmetry constraints imposed by the preserved $C_{nz}$ [=$\exp(-\frac{i\pi}{n}\sigma_z)$], $PT$ [$=\tau_x\otimes i\sigma_y K$], and $M_xT$ [$=-i\sigma_z K$] symmetries, the most general symmetry-allowed coupling Hamiltonian can be obtained as
\begin{equation}
\tag{S10}
    H_{\mbox{\tiny{coup}}}=(\Delta-Bk^2)\tau_x\otimes\sigma_0+R_1\frac{k_+^n-k_-^n}{2i}\tau_y\otimes\sigma_0+R_3\frac{k_+^n+k_-^n}{2}\tau_x\otimes\sigma_0
\end{equation}

\begin{figure}[htbp]
\centering
\includegraphics[width=0.5\textwidth]{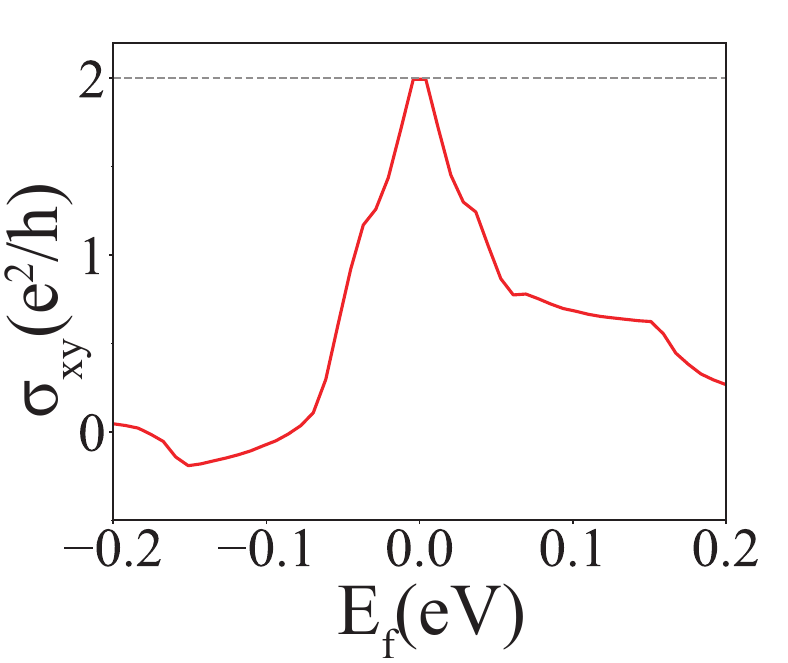}
\caption{\textbf{Anomalous Hall conductivity for the $\mathbf{n=2}$ case with the neglected $\mathbf{R_3}$ coupling term when $\mathbf{u_-<u<u_+}$.} The parameters are chosen as $u=0.1\ \mathrm{eV},m=0.05\ \mathrm{eV}, v=1\ \mathrm{eV\cdot \AA}, \Delta=0.05\ \mathrm{eV},B=5\ \mathrm{eV\cdot \AA^2}, R_1=10\ \mathrm{eV\cdot\AA^2},R_2=10\ \mathrm{eV\cdot\AA^2},R_3=2\ \mathrm{eV\cdot\AA^2}$ in the above numerical calculations.}
\label{fig1}
\end{figure}

Apart from the first two terms discussed in the main text, the third higher-order coupling $R_3$ term is also symmetry-allowed which is neglected in the main text for simplicity. Here, we will show that it will not affect the main results of our work. We take the $n=2$ case with the full Hamiltonian written as
\begin{equation}
\tag{S11}
H= m(\mathbf{k})\tau_z\otimes\sigma_z+u\tau_z\otimes\sigma_0+v\tau_z\otimes(k_x\sigma_y-k_y\sigma_x)+\Delta(\mathbf{k})\tau_x\otimes\sigma_0+R_1g_2(k_x,k_y)\tau_y\otimes\sigma_0,
\end{equation}
where $m(\mathbf{k})=m+R_2(k_x^2-k_y^2)$, and $\Delta(\mathbf{k})=\Delta-Bk^2+R_3(k_x^2-k_y^2)$. Through a similar procedure as above, we can find the critical potential and the intertwined Dirac cones as
\begin{equation}
\tag{S12}
    \left\{\begin{aligned}
    & u_+\equiv u_j=\sqrt{(m+\frac{\Delta R_2}{B-R_3})^2+\frac{v^2\Delta}{B-R_3}}\\
    & k_j=\sqrt{\frac{\Delta}{B-R_3}}\\
    & \theta_j=j\pi/2,\quad j=0,2
    \end{aligned}\right.
\end{equation}
and
\begin{equation}
\tag{S13}
    \left\{\begin{aligned}
    & u_-\equiv u_j=\sqrt{(m-\frac{\Delta R_2}{B+R_3})^2+\frac{v^2\Delta}{B+R_3}}\\
    & k_j=\sqrt{\frac{\Delta}{B+R_3}}\\
    & \theta_j=j\pi/2,\quad j=1,3
    \end{aligned}\right.
\end{equation}

As we can see, if $R_3<B$ is satisfied, there are still two successive gap-closing processes at $u_-$ and $u_+$, respectively, and the calculation of anomalous Hall conductivity (AHC) [shown in Fig. S\ref{fig1}] confirms that there is a high-Chern-number phase ($|C|=2$) between $u_-$ and $u_+$. However, due to the nonzero $R_3$ term, the gap-closing points are no longer located on the circle with radius $k_0=\sqrt{\Delta/B}$. In contrast, if $R_3>B$, $k_j=\sqrt{\Delta/(B-R_3)}$ with $j=0,2$ will not exist, indicating that there will not be another gap-closing process occurring at $u=u_+$. As a result, there exists an upper critical value of $R_3$ for the emergence of the intermediate high-Chern-number state. Nevertheless, the $R_3$ coupling term does not fundamentally affect the results in the main text and is thus reasonably neglected for simplicity.

\section{Berry curvatures and anomalous Hall conductivities of model with $\mathbf{n=2,4,6}$}
\begin{figure}[htbp]
\centering
\includegraphics[width=0.8\textwidth]{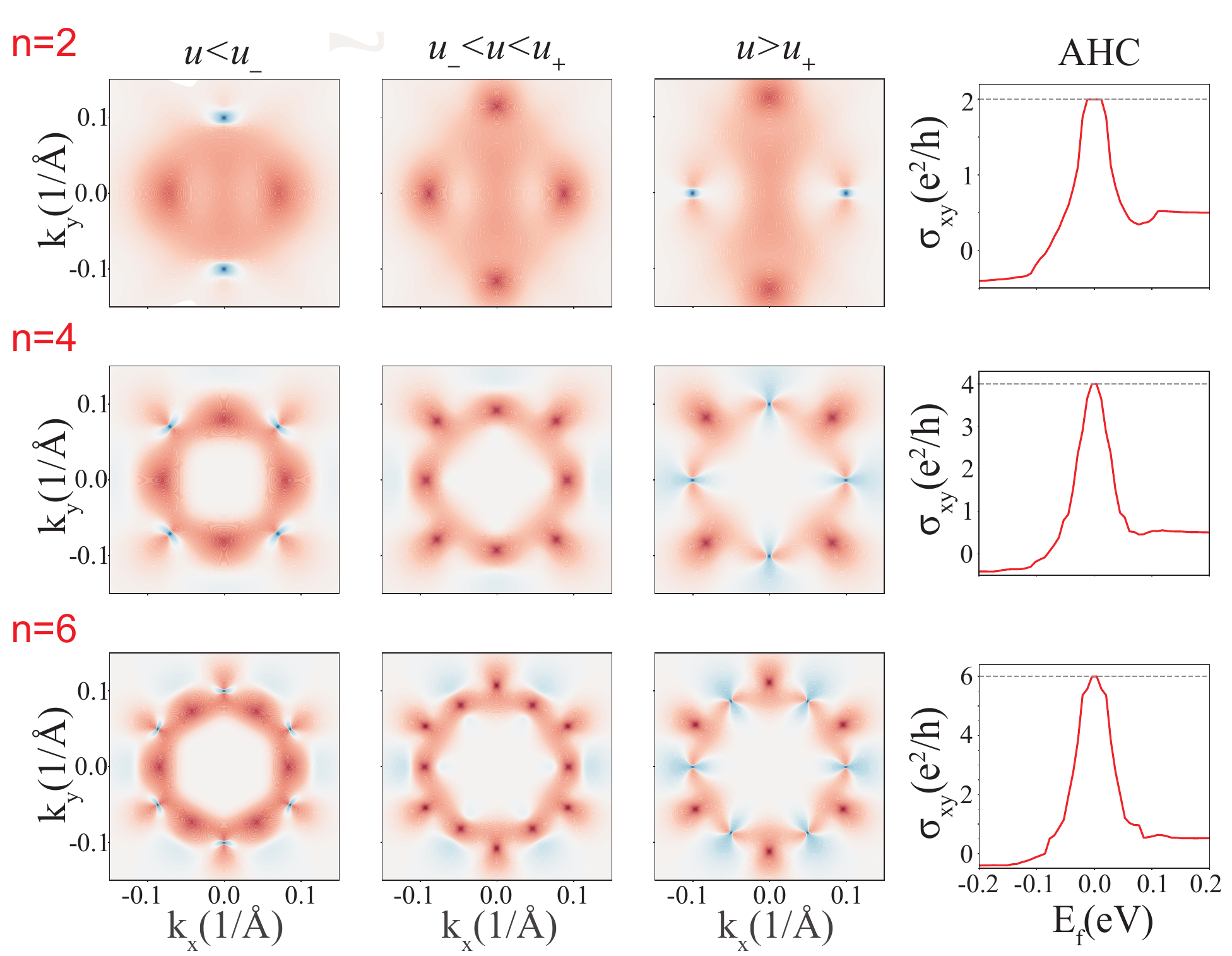}
\caption{\textbf{Berry curvatures and anomalous Hall conductivities of model with $\mathbf{n=2,4,6}$.} The parameters are chosen as $m=0.05\ \mathrm{eV}, v=1\ \mathrm{eV\cdot \AA}, \Delta=0.05\ \mathrm{eV},B=5\ \mathrm{eV\cdot \AA^2}, R_1=10^{n-1}\ \mathrm{eV\cdot\AA}^{n},R_2=10^{n-1}\ \mathrm{eV\cdot\AA}^{n}$ in the above numerical calculations.}
\label{fig2}
\end{figure}

In addition to the $n=3$ case in the main text, here, we show the numerical calculations of the Berry curvatures and anomalous Hall conductivities for the $n=2$ (first row), $n=4$ (second row), and $n=6$ (third row) cases in Fig. S\ref{fig2}.

\section{The magnetic ground state of $\rm \mathbf{MnBi_2Te_4/(Bi_2Te_3)_m/MnBi_2Te_4}$ ($\rm \mathbf{m=0,1,2}$)}
Here, we only consider the magnetic order with Mn atoms' magnetic moments towards the out-of-plane direction. We compare the total energy per cell of $\rm MnBi_2Te_4/(Bi_2Te_3)_m/MnBi_2Te_4$ ($\rm m=0,1,2$) with ferromagnetic (FM) order and antiferromagnetic (AFM) order under zero and the typical values of electric field in the calculations. The results are listed in Table S\ref{tab1} and Table S\ref{tab2}. As shown in Table S\ref{tab1} (Table S\ref{tab2}), for m=0,1,2, the total energy per cell of the AFM order is lower than FM order of 1.46 (1.11) meV, 0.19 (5.41) meV and 0.05 (2.97) meV under zero (the typical values of) eletric field. These results verify the magnetic ground state of $\rm MnBi_2Te_4/(Bi_2Te_3)_m/MnBi_2Te_4$ ($\rm m=0,1,2$) is AFM order state in the range of elelctric field of the calculations. 

\begin{table}[htbp]
\renewcommand{\thetable}{1}
\centering
\begin{tabular}{|c|c|c|} 
 \hline
 m & Total energy of FM order & Total energy of AFM order \\ 
 \hline
 0 & 0 meV & -1.46 meV \\ 
 \hline
 1 & 0 meV & -0.19 meV \\
 \hline
 2 & 0 meV & -0.05 meV \\ 
 \hline
\end{tabular}
\caption{\textbf{Total energy per cell of $\rm \mathbf{MnBi_2Te_4/(Bi_2Te_3)_m/MnBi_2Te_4(m=0,1,2)}$ with out-of-plane FM and AFM magnetic order.} The total energy of the FM order has been set to zero.}
\label{tab1}
\end{table}

\begin{table}[htbp]
\renewcommand{\thetable}{2}
\centering
\begin{tabular}{|c|c|c|c|} 
 \hline
 m & Electric field & Total energy of FM order & Total energy of AFM order \\ 
 \hline
 0 & 0.015 $\mathrm{eV/\AA}$ & 0 meV & -1.12 meV \\ 
 \hline
 1 & 0.015 $\mathrm{eV/\AA}$ & 0 meV & -5.41 meV \\
 \hline
 2 & 0.010 $\mathrm{eV/\AA}$ & 0 meV & -2.97 meV \\ 
 \hline
\end{tabular}
\caption{\textbf{Total energy per cell of $\rm \mathbf{MnBi_2Te_4/(Bi_2Te_3)_m/MnBi_2Te_4(m=0,1,2)}$ with out-of-plane FM and AFM magnetic order under the typical values of electric field.} The total energy of the FM order has been set to zero.}
\label{tab2}
\end{table}

\section{Wilson-loops of $\rm \mathbf{MnBi_2Te_4/Bi_2Te_3/MnBi_2Te_4}$}
\begin{figure}[htbp]
\centering
\includegraphics[width=0.8\textwidth]{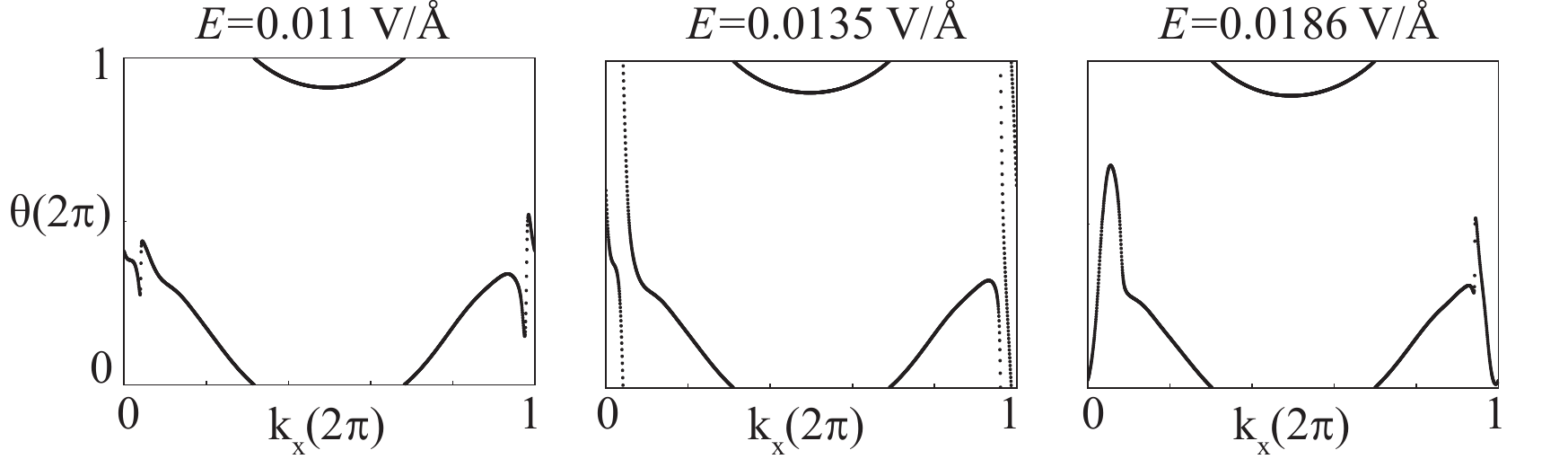}
\caption{\textbf{Wilson-loops of $\rm \mathbf{MnBi_2Te_4/(Bi_2Te_3)/MnBi_2Te_4}$ under different electric fields.} The electric fields are $E=0.011\ \mathrm{V/\AA}$,$E=0.0135\ \mathrm{V/\AA}$, and $E=0.0186\ \mathrm{V/\AA}$.}
\label{fig3}
\end{figure}

Here, we show the calculation of the Chern number of $\rm MnBi_2Te_4/Bi_2Te_3/MnBi_2Te_4$ by Wilson-loop methods (the Wilson-loop winding number equals to the Chern number~\cite{wilson-loop}) under different electric field values in Fig. S\ref{fig3}. As shown in Fig. S\ref{fig3}, when $E=0.011\ \mathrm{V/\AA}$, the Wilson-loop winding number is 0, when $E=0.0135\ \mathrm{V/\AA}$, the Wilson-loop winding number is 3, and when $E=0.0186\ \mathrm{V/\AA}$, the Wilson-loop winding number returns to 0.

\begin{figure}[htbp]
\centering
\includegraphics[width=0.8\textwidth]{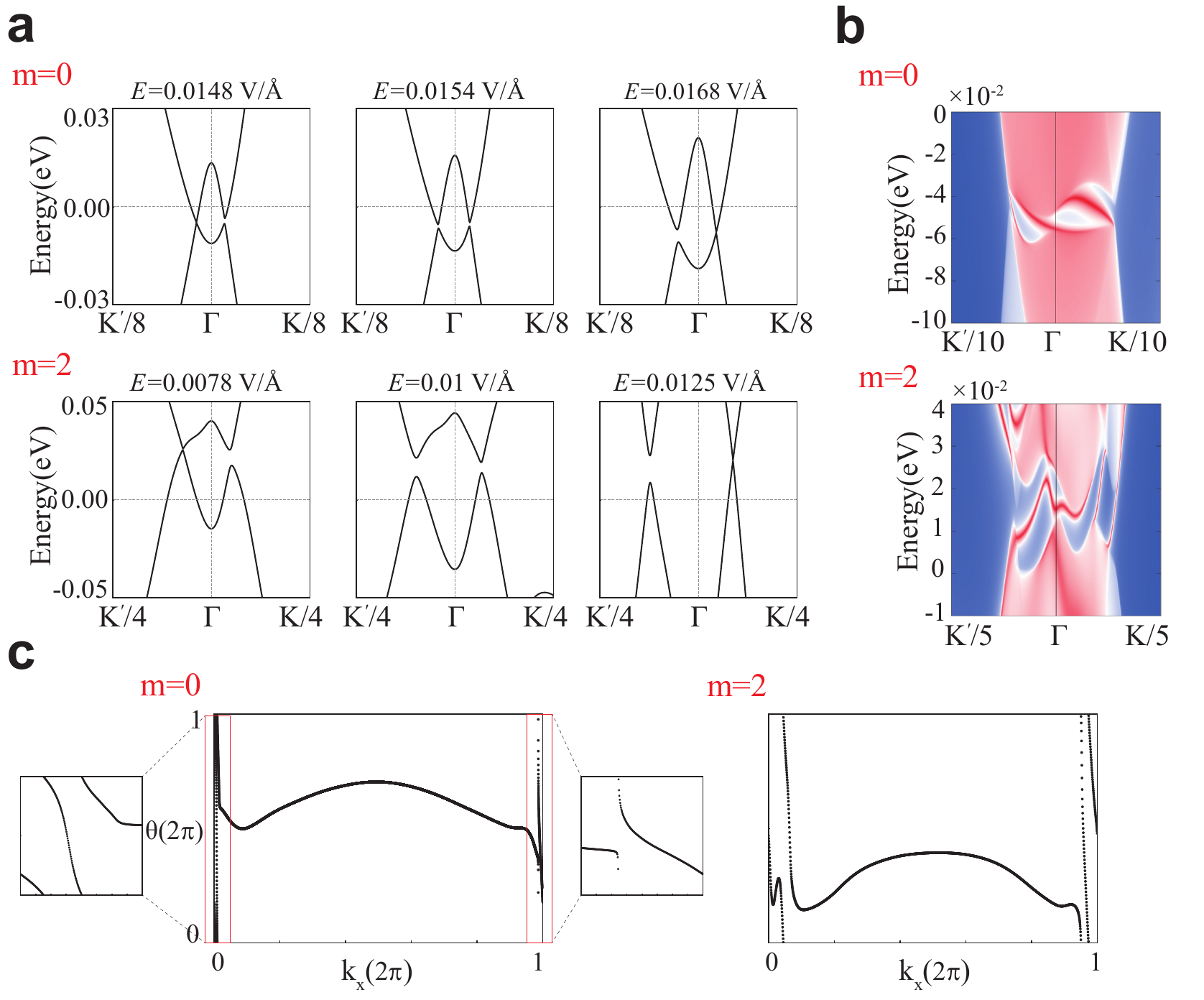}
\caption{\textbf{Band structures, Wilson-loops, and edge states of $\rm \mathbf{MnBi_2Te_4/(Bi_2Te_3)_m/MnBi_2Te_4}$ ($\rm \mathbf{m=0,2}$).} (\textbf{a}) Band structures of $\rm MnBi_2Te_4/(Bi_2Te_3)_m/MnBi_2Te_4$ ($\rm m=0,2$) under diffrent electric fields: $E=0.0148\ \mathrm{V/\AA}$,$E=0.0154\ \mathrm{V/\AA}$, $E=0.0168\ \mathrm{V/\AA}$ for $\mathrm{m}=0$; $E=0.0078\ \mathrm{V/\AA}$,$E=0.01\ \mathrm{V/\AA}$, $E=0.0125\ \mathrm{V/\AA}$ for $\mathrm{m}=2$. (\textbf{b}) Edge states of $\rm MnBi_2Te_4/(Bi_2Te_3)_m/MnBi_2Te_4$ ($\rm m=0,2$) when $E=0.0162\ \mathrm{V/\AA}$ ($\mathrm{m}=0$) and $E=0.01\ \mathrm{V/\AA}$ ($\mathrm{m}=2$). (\textbf{c}) Wilson-loops of $\rm MnBi_2Te_4/(Bi_2Te_3)_m/MnBi_2Te_4$ ($\rm m=0,2$) when $E=0.0162\ \mathrm{V/\AA}$ ($\mathrm{m}=0$) and $E=0.01\ \mathrm{V/\AA}$ ($\mathrm{m}=2$).}
\label{fig4}
\end{figure}

\section{Band structures, Wilson-loops, and edge states of $\rm \mathbf{MnBi_2Te_4/(Bi_2Te_3)_m/MnBi_2Te_4}$ ($\rm \mathbf{m=0,2}$)}
Here, we show that band structures, Wilson-loops, and edge states of $\rm MnBi_2Te_4/(Bi_2Te_3)_m/MnBi_2Te_4$ ($\rm \mathrm{m}=0,2$) in Fig. S\ref{fig4} to exhibit the 0-3-0 transition of Chern number of them. Fig. S\ref{fig4}a shows band structures of $\rm MnBi_2Te_4/(Bi_2Te_3)_m/MnBi_2Te_4$ ($\rm \mathrm{m}=0,2$) under different electric fields. For the $\mathrm{m}=0$ ($\mathrm{m}=2$) case, the energy gap closes at $E=0.0148\ \mathrm{V/\AA}$ ($E=0.0078\ \mathrm{V/\AA}$) and reopens when the electric field continues to increase, and when $E=0.0168\ \mathrm{V/\AA}$ ($E=0.0125\ \mathrm{V/\AA}$), the gap closes again. Fig. S\ref{fig4}c shows the calculation of the Wilson-loops for the $\mathrm{m}=0$ and $\mathrm{m}=2$ cases when $E=0.0162\ \mathrm{V/\AA}$ and $E=0.01\ \mathrm{V/\AA}$, respectively, both of which exhibit a Wilson-loop winding number of 3. It confirms the existence of a high-Chern-number phase with $C=3$. Fig. S\ref{fig4}b shows the edge states of $\mathrm{m}=0$ and $\mathrm{m}=2$ case when $E=0.0162\ \mathrm{V/\AA}$ and $E=0.01\ \mathrm{V/\AA}$ respectively, verifying the Chern number is indeed 3.

\bibliography{supref}